\documentclass[conference]{IEEEtran}
\IEEEoverridecommandlockouts
\usepackage{cite}
\usepackage{amsmath,amssymb,amsfonts}
\usepackage{algorithmic}
\usepackage{graphicx}
\usepackage{textcomp}
\usepackage{booktabs}
\usepackage{xcolor}
\def\BibTeX{{\rm B\kern-.05em{\sc i\kern-.025em b}\kern-.08em
    T\kern-.1667em\lower.7ex\hbox{E}\kern-.125emX}}

\begin{document}

\title{Survey on Token-Based Distributed Mutual Exclusion Algorithms}

\author{
    \IEEEauthorblockN{Elahe Tohidi\IEEEauthorrefmark{1},
                      Seyed Sattar Lotfi Fatemi \IEEEauthorrefmark{2}}
    \IEEEauthorblockA{\IEEEauthorrefmark{1}Department of Mathematical Sciences, Sharif University of Technology, Tehran, Iran\\
                      Email: elahetohidi2017@gmail.com}
    \IEEEauthorblockA{\IEEEauthorrefmark{2}Department of Mathematical Sciences, Sharif University of Technology, Tehran, Iran\\
                      Email: Sat.fate@sharif.edu}
}

\maketitle

\begin{abstract}
In large-scale distributed environments, avoiding concurrent access to the same resource by multiple processes becomes a core challenge, commonly termed \emph{distributed mutual exclusion} (DME). Token-based mechanisms have long been recognized as an effective strategy, wherein a solitary \emph{token} is handed around among processes as the key that allows access to the critical section. By doing so, they often reduce the messaging overhead compared to alternate methods.

This work surveys the significance of mutual exclusion in distributed computing and examines \emph{token-based} solutions across various network models (including tree-based, ring-based, fully interconnected graphs, mesh structures, and ad hoc networks). We also delve into essential performance measures such as communication costs and strategies for fault tolerance, then branch into specialized variants, such as \emph{k-mutual exclusion} and \emph{self-stabilizing} algorithms. Furthermore, a specialized approach that relies on \emph{finite projective planes} is introduced to highlight how certain protocols can perform efficiently under both best- and worst-case conditions. Lastly, we explore future directions involving machine learning for token predictive routing and blockchain techniques to resist adversarial behavior. This aims to provide a thorough yet accessible overview of token-based DME approaches, together with insights on emerging research trends.
\end{abstract}


\section{Introduction}
\label{sec:intro}
Distributed systems often comprise numerous autonomous computers collaborating to share data, hardware, or other resources \cite{tanenbaum2013distributed}. A key principle in these systems is known as \emph{mutual exclusion}, ensuring that only one participant at a time engages in a critical section where shared resources are manipulated. Enforcing this principle across multiple machines without a unified clock or common memory is the essence of the \emph{Distributed Mutual Exclusion (DME)} problem.

In general, the literature distinguishes DME solutions into two primary categories:
\begin{enumerate}
    \item \textbf{Permission-based (non-token) algorithms}, such as the Ricart--Agrawala scheme \cite{ricart1981optimal}, where a node must acquire authorization from others (or a designated quorum) before using the critical resource.
    \item \textbf{Token-based algorithms}, exemplified by Suzuki--Kasami \cite{suzuki1985distributed}, where a specific token grants unique access rights to the critical section.
\end{enumerate}

We concentrate on \emph{token-based} methods. Section~\ref{sec:foundations} covers the fundamental requirements for mutual exclusion, and Section~\ref{sec:token_classification} outlines representative token-based protocols that employ various topological assumptions. In Section~\ref{sec:analysis_comparisons}, we compare these algorithms regarding performance indices (e.g., message complexity) and fault resilience, plus examine how they differ from permission-based approaches. Advanced topics such as group-based resource sharing and self-stabilization are addressed in Section~\ref{sec:adv_extensions}. Section~\ref{sec:open_issues} identifies current challenges, including the rising need for robust solutions in highly mobile networks, opportunities to leverage machine learning, and employing blockchain-inspired security. Meanwhile, Section~\ref{sec:finite_pp} illustrates how an alternate construction based on \emph{finite projective planes} can attain $O(1)$ best-case complexity and $O(\sqrt{N})$ in the worst case. The final discussion in Section~\ref{sec:conclusion} reflects on the overall state of token-based DME and where it may develop in coming years.

\section{Fundamental Principles of Mutual Exclusion}
\label{sec:foundations}

\subsection{Basic Requirements}
Any valid DME solution enforces:
\begin{itemize}
    \item \textbf{Safety}: Strictly one process occupies the critical section at a given moment.
    \item \textbf{Liveness}: Each process that requests the critical section is eventually allowed to proceed (i.e., no infinite blocking).
    \item \textbf{Fairness}: Typically, the order in which requests are served follows a queue-based or first-come-first-served manner, though some protocols relax fairness to minimize overhead.
\end{itemize}

\subsection{Token-based vs.\ Permission-based Approaches}
\label{subsec:token_vs_permission}
Permission-based solutions, including Ricart--Agrawala \cite{ricart1981optimal}, require that processes explicitly receive responses from other nodes before entering the critical section. In contrast, \textbf{token-based} models permit exclusive access to whichever participant holds a logically unique token, often cutting down on the communication required once a token is in transit. Nevertheless, designers must guard against situations where the token is mistakenly duplicated or dropped.

\subsection{Performance Metrics}
Key measures include:
\begin{itemize}
    \item \textbf{Message Complexity}: The volume of message exchanges needed for a process to enter and exit the critical section.
    \item \textbf{Synchronization Delay}: The gap (in rounds or hops) between a process releasing the critical section and the next process gaining it.
    \item \textbf{Response Time}: How long a requesting node waits before proceeding with its critical section.
    \item \textbf{Fault Tolerance}: The ability of an algorithm to stay correct or recover despite node or link failures, or other disruptions.
\end{itemize}

\subsection{Historical Background}
Foundational work on synchronization in distributed systems emerged in the 1970s, with Lamport proposing a logical clock-based ordering \cite{lamport1978time}. Subsequent developments, like the Ricart--Agrawala scheme \cite{ricart1981optimal}, provided permission-based solutions. Suzuki and Kasami’s token-based strategy \cite{suzuki1985distributed} was especially notable for slashing the messaging overhead in a fully connected topology. Researchers have since adapted token-based designs for rings, trees, grids, and mobile networks, alongside emphasizing reliability, multiple-token usage, and self-stabilization techniques \cite{parihar2021token}.

\section{Token-Based Algorithms by Network Topology}
\label{sec:token_classification}
Token-based methods commonly incorporate assumptions about the network layout, ranging from fully connected nodes to ring or tree formations, mesh configurations, and more fluid ad hoc environments \cite{parihar2021token}.

\subsection{Fully Connected Networks}
\subsubsection{Suzuki--Kasami Algorithm \cite{suzuki1985distributed}}
A process broadcasts a \texttt{REQUEST} whenever it wants to enter the critical section. The current token holder forwards the token upon noticing an updated request index. Though $O(N)$ messages may be used per request, this can still be an improvement over naive permission-based setups needing $2(N-1)$ messages.

\subsubsection{Singhal’s Heuristic \cite{singhal1989heuristic}}
Instead of broadcasting widely, each node identifies a narrower subset of likely token holders, lowering the number of messages. This is especially beneficial with frequent critical-section access.

\subsubsection{Fault-Tolerant Variants \cite{mishra1990fault,nishio1990resilient}}
Some designs account for lost or crashed token owners, regenerating tokens with incrementing epoch numbers and invalidating any residual tokens from older epochs.

\subsection{Tree-Based Algorithms}
Token movement can be contained within a spanning tree that covers all nodes.

\subsubsection{Raymond’s Tree Algorithm \cite{raymond1989tree}}
Messages propagate along the tree toward the node believed to hold the token. If the tree remains balanced, each request can be served in around $O(\log N)$ messages.

\subsubsection{Naimi--Trehel Path Reversal \cite{parihar2021token}}
Pointers are altered as requests travel, effectively bringing the token closer to nodes that request it most frequently, which can be advantageous under skewed demand.

\subsubsection{Failure Handling}
If a crucial link or node in the tree fails, local routes must be mended or restructured so that token access remains feasible.

\subsection{Ring-Based Protocols}
\subsubsection{Simple Token Ring}
Here, nodes form a cycle, and the token goes around in a loop. Under steady demand, each node obtains the token in a fixed order, but worst-case waiting can be $O(N)$ if a node just misses the token.

\subsubsection{Chordal/Shortcut Rings \cite{parihar2021token}}
By adding shortcut links, token movement speeds up at the expense of introducing extra rules to ensure the token remains unique and loop-free.

\subsection{Mesh or 2D Grid Topologies}
In a two-dimensional grid, nodes typically connect to adjacent neighbors (e.g., up, down, left, right). Strategies might split request routing along one axis and token movement along another, yielding complexities between $O(\sqrt{N})$ and $O(N)$ depending on how effectively traffic is managed.

\subsection{Dynamic and Ad Hoc Networks}
\label{subsec:dynamic}
Ad hoc environments such as MANETs or FANETs can see frequent topological reconfigurations \cite{khanna2019mrme,khanna2020local,walter2001mutual}.

\subsubsection{Walter et al.\ \cite{walter2001mutual}}
Maintains a Directed Acyclic Graph (DAG), adjusting edges as nodes move to avoid cycles and ensure a singular path to the token holder.

\subsubsection{FANET Approaches (MRME, RCLME) \cite{khanna2019mrme,khanna2020local}}
\begin{itemize}
    \item \emph{MRME}: Retains a standby token copy to quickly recreate it if the primary holder vanishes due to mobility issues.
    \item \emph{RCLME}: Deploys fuzzy logic to elect local leaders in UAV swarms, thereby restricting token transmissions to smaller clusters.
\end{itemize}

\section{Performance Analysis, Fault Tolerance, and Comparison}
\label{sec:analysis_comparisons}
\subsection{Message Complexity and Synchronization Delay}
\begin{itemize}
    \item \textbf{Fully Connected}: Algorithms like Suzuki--Kasami entail about $O(N)$ messages per request, typically superior to simpler permission-based approaches.
    \item \textbf{Tree-Based}: Techniques such as Raymond’s can maintain about $O(\log N)$ on balanced trees but degrade if the tree is skewed.
    \item \textbf{Ring-Based}: Straightforward, but a process can still wait $O(N)$ cycles for the token.
    \item \textbf{Mesh/2D}: Usually between $O(\sqrt{N})$ and $O(N)$, shaped by how requests and token routes are organized.
    \item \textbf{Dynamic/Ad Hoc}: Overheads are highly variable since nodes may come and go. Clever local caching or adaptivity mitigates cost but cannot always avoid adverse cases.
\end{itemize}

\subsection{Fault-Tolerance Mechanisms}
\begin{enumerate}
    \item \textbf{Node Crash}: A timeout-based strategy identifies missing token holders and regenerates a valid token with a new version number.
    \item \textbf{Link Failure}: In tree or ring topologies, reconfigurations or link bypassing are required; fully connected graphs can more freely choose alternate links.
    \item \textbf{Token Duplication or Loss}: Typically, sequence identifiers or logs discredit outdated tokens in favor of newly regenerated ones.
\end{enumerate}

\subsection{Comparison with Permission-Based Methods}
\label{subsec:compare_permission_based}
Permission-based methods like Ricart--Agrawala may see $2(N-1)$ messages for each critical-section request, though quorum-driven protocols (e.g., Maekawa) can reduce this to around $O(\sqrt{N})$. Token-based systems usually do well under frequent usage scenarios, as the token migrates to the next requester. Nonetheless, extra caution is needed around potential token misplacements. Ultimately, which approach is optimal depends on the network’s scale, request distribution, and tolerance for partial failures.

\section{Advanced Extensions: Group/k-Mutual Exclusion and Self-Stabilization}
\label{sec:adv_extensions}

\subsection{Group/k-Mutual Exclusion}
When multiple processes (up to \emph{k}) can safely use a resource simultaneously, \emph{k} tokens or partial tokens must be circulated.

\subsubsection{Multi-Token Raymond}
A direct extension of Raymond’s approach that handles multiple tokens concurrently, carefully preventing conflicts when multiple processes request the resource.

\subsubsection{Other Approaches}
Ring- and fully connected-based algorithms can likewise be augmented by tracking how many tokens are in use and stopping new ones from being issued once \emph{k} tokens have been distributed.

\subsection{Self-Stabilizing DME}
\textbf{Self-stabilization} \cite{dijkstra1974self} indicates the algorithm will return to a valid single-token state from any erroneous configuration.

\subsubsection{Kiniwa’s BFS Tree \cite{kiniwa2006request}}
Nodes create a BFS tree aimed at the (suspected) token holder. If multiple tokens arise, collisions resolve so that only one legitimate token remains.

\subsubsection{Chen \& Welch for MANETs \cite{chen2005self}}
Uses two distinct tokens to handle group membership and mutual exclusion. Older tokens become invalid once they meet a newer epoch version.

\section{An Additional Token-Based Method Using Finite Projective Planes}
\label{sec:finite_pp}
Aside from typical layouts (trees, rings, etc.), a unique approach employs \emph{finite projective planes} to obtain an $O(\sqrt{N})$ worst-case overhead and an $O(1)$ best case.

\subsection{High-Level Idea}
Nodes are organized into \textbf{superior} and \textbf{inferior} sets through a projective plane arrangement, providing:
\begin{itemize}
    \item A single node in common for every pair of arbitrator sets, removing repeated or overly broad requests.
    \item Direct knowledge by each node’s superior set on who holds the token.
    \item A single physical token, to ensure that local checks confirm no inferior node currently has it before requesting.
\end{itemize}

\subsection{Algorithm Outline}
\begin{enumerate}
    \item \textbf{Initialization}: One node starts with the token, notifying its \emph{superior} group.
    \item \textbf{Requesting the Token}: The requester checks its \emph{inferior} set for token possession, then asks its superiors if the token is elsewhere.
    \item \textbf{Token Passing}: A node holding the token grants it if requested or propagates the request to whichever node is indicated by the projective-plane structure.
    \item \textbf{Exiting the CS}: On leaving the critical section, the node updates superiors and, if another request is waiting, hands over the token.
\end{enumerate}

\subsection{Correctness and Performance}
\begin{itemize}
    \item \textbf{Mutual Exclusion}: A single valid token is needed to enter the critical section, preventing concurrency issues.
    \item \textbf{Starvation Freedom}: Requests eventually reach the token, thanks to a structured approach ensuring at most one intersection node for each pair of sets.
    \item \textbf{Deadlock Avoidance}: There is always a path from any requester to the actual token holder.
    \item \textbf{Message Complexity}: In the best scenario, no network-level searching is needed ($O(1)$). In the worst case, requesting nodes may have to consult about $O(\sqrt{N})$ sets.
\end{itemize}

This method can outperform some consensus-based strategies (like Maekawa’s) in scenarios where repeated requests come from specific nodes.

\section{Open Challenges and Research Directions}
\label{sec:open_issues}

\subsection{Highly Dynamic Networks (FANET, VANET, MANET)}
Frequent disconnections and reconnections can lead to lost or duplicate tokens:
\begin{itemize}
    \item \textbf{Partition-Aware Methods}: Retain distinct tokens for separate partitions; safely unify these tokens upon merging networks.
    \item \textbf{Energy Minimization}: In UAV or sensor setups, communication costs must be carefully limited.
\end{itemize}

\subsection{Machine Learning Integration}
Predictive techniques can streamline token handling:
\begin{itemize}
    \item \textbf{Request Forecasting}: Position the token closer to nodes that are likely to require the critical section soon.
    \item \textbf{Adaptive Timeout}: Use learning algorithms to tune timeouts, lowering the chance of prematurely concluding the token is lost.
\end{itemize}

\subsection{Blockchain-Level Security}
If nodes are malicious, they might forge or misuse tokens:
\begin{itemize}
    \item \textbf{Immutable Transfer Logs}: Employ ledger-style records for each token handoff so that inconsistencies stand out.
    \item \textbf{Lightweight Consensus}: Byzantine fault-tolerant or proof-of-authority mechanisms can defend token legitimacy in partially trusted networks.
\end{itemize}

\subsection{Data Centers and Edge/Fog Computing}
Token-based approaches remain relevant in large-scale distributed settings:
\begin{itemize}
    \item \textbf{Hierarchical Organization}: Small clusters each use local tokens, with a global coordinator overseeing cross-cluster requests.
    \item \textbf{Scalable Trees/Projective Planes}: Balanced structures for thousands of nodes can maintain manageable overhead.
\end{itemize}

\subsection{Combining Token- and Permission-Based Methods}
Certain hybrid solutions leverage token-based concurrency within localized groups, but rely on permission or quorum checks for broader system-level coordination.

\subsection{Energy Constraints}
Especially critical in battery-powered contexts:
\begin{itemize}
    \item \textbf{Directed Token Passing}: Only pass tokens to nodes that have declared a need, skipping idle participants.
    \item \textbf{Sleep Schedules}: Power down nodes lacking active requests until they must handle or request the token.
\end{itemize}

\subsection{Large-Scale Self-Stabilization}
While self-stabilizing protocols can cope with arbitrary initial faults, overhead may rise significantly for big or highly mobile systems. Ongoing research focuses on localized repair that avoids major performance penalties.

\section{Conclusion and Unified Perspectives}
\label{sec:conclusion}

In many distributed scenarios, \textbf{token-based mutual exclusion} remains an effective tactic for governing access to shared resources. Compared to permission-based strategies that hinge on multi-party acknowledgments, passing a single authoritative token can be more communication-efficient if the token is sensibly routed.

\begin{itemize}
    \item \textbf{Influence of Network Topology}: Fully connected setups lend themselves to simple broadcast-then-token-forward methods (e.g., Suzuki--Kasami), while tree-based or ring-based structures offer different trade-offs in terms of messaging or wait times.
    \item \textbf{Finite Projective Planes Approach}: By structuring nodes into superior/inferior sets, one can achieve $O(1)$ best-case messaging and $O(\sqrt{N})$ in worst cases—especially helpful if a handful of processes frequently request the resource.
    \item \textbf{Mobility and Fault Tolerance}: Techniques must handle token loss, duplication, or node/link failures without breaking mutual exclusion.
    \item \textbf{Promising Enhancements}: Group/k-mutual exclusion, self-stabilization, machine learning–driven token forwarding, and blockchain-level defenses show how token-based schemes adapt to modern computational demands, including UAV fleets and huge data centers.
\end{itemize}

Looking ahead, token-based DME algorithms offer a robust, evolving solution path. Whether augmented by specialized data structures, predictive intelligence, or secure handover records, token passing will likely continue to be a foundational mechanism for coordination in distributed systems of all scales and complexities.

\vspace{1em}

\bibliographystyle{IEEEtran}

\end{document}